\documentstyle[amsfonts,aps,twocolumn,pra,psfig]{revtex}

\begin{document}
\title{Temperature dependence of the Casimir effect between metallic mirrors}
\author{Cyriaque Genet, Astrid Lambrecht and Serge Reynaud}
\address{Laboratoire Kastler Brossel \thanks{
Unit\'{e} de l'Ecole Normale Sup\'{e}rieure, de l'Universit\'{e} Pierre
et Marie Curie, et du Centre National de la Recherche Scientifique.}, \\
Campus Jussieu, case 74, \\
75252 Paris Cedex 05, France}
\date{February 22, 2000}
\maketitle

\begin{abstract}
We calculate the Casimir force and free energy for plane metallic mirrors at
non-zero temperature. Numerical evaluations are given with temperature and
conductivity effects treated simultaneously. The results are compared with 
the approximation where both effects are treated independently and the
corrections simply multiplied. The deviation between the exact and
approximated results takes the form of a temperature dependent function for
which an analytical expression is given. The knowledge of this function
allows simple and accurate estimations at the \% level.
\end{abstract}

\section{Introduction}

The Casimir force \cite{Casimir48} has been observed in a number of 
`historic' experiments \cite{Deriagin57,Sparnaay,Tabor68,Sabisky73}. 
It has been measured recently with an improved experimental precision 
\cite{Lamoreaux97,Mohideen98,Roy99}. This should allow for an accurate 
comparison with the predictions of Quantum Field Theory, provided that 
these predictions account for the differences between real
experiments and the idealized Casimir situation. In particular, experiments
are performed at room temperature between metallic mirrors and not at zero
temperature between perfect reflectors. The theoretical expectations should be
computed with a high accuracy if the aim is to test agreement between theory
and experiment at, say, the 1\% level. The efforts for accuracy are also
worth for making it possible to control the effect of Casimir force when
studying small short range forces \cite{Fischbach,Carugno97,Bordag99}.

The influence of thermal field fluctuations on the Casimir force are known
to become important for distances of the order of a typical length 
\cite{Lifshitz,Mehra67,Brown69,Schwinger78} 
\begin{equation}
\lambda _{\rm T}=\frac{2\pi c}{\omega _{\rm T}}=\frac{\hbar c}{k_{B}T}
\label{lambdaT}
\end{equation}
When evaluated at room temperature, this length $\lambda _{\rm T}$ is
approximately $7\mu $m. In contrast, the finite conductivity of metals has
an appreciable effect for distances smaller than or of the order of the
plasma wavelength $\lambda _{\rm P}$ determined by the plasma frequency $%
\omega _{\rm P}$ of the metal (see \cite{Lambrecht00} and references
therein) 
\begin{equation}
\lambda _{\rm P}=\frac{2\pi c}{\omega _{\rm P}}  \label{lambdaP}
\end{equation}
For metals used in the recent experiments, this wavelength lies in the range
0.1$\mu $m-0.2$\mu $m. This means that conductivity and thermal corrections
to the Casimir force are important in quite different distance ranges.
Thermal corrections are usually ignored in the sub-$\mu $m range where the
effect of imperfect reflection is significant whereas the conductivity
correction is unimportant above a few $\mu $m where the effect of temperature
becomes appreciable. This explains why the 2 corrections are usually treated
independently from each other. When an accurate comparison between
experimental and theoretical values of the Casimir force is aimed at, 
the error induced by this approximation has however to be precisely 
evaluated. Furthermore, the region of overlap of the two
corrections is precisely in the $\mu $m range, which is a crucial distance
range for the comparison between experiment and theory.

The purpose of this paper is to give an accurate evaluation of the Casimir
force $F$ taking into account finite conductivity and temperature corrections at
the same time. To characterize the whole correction, we will compute the
factor $\eta _{\rm F}$ describing the combined effect of conductivity and
temperature 
\begin{eqnarray}
\eta _{\rm F} &=&\frac{F}{F_{\rm Cas}}  \nonumber \\
F_{\rm Cas} &=&\frac{\hbar cA\pi ^{2}}{240L^{4}}  
\label{Fcasimir}
\end{eqnarray}
$F_{\rm Cas}$ is the ideal Casimir force corresponding to perfect mirrors
in vacuum. $L$ is the distance between the mirrors, $A$ their surface and $%
\hbar $ and $c$ respectively the Planck constant and the speed of light. We
will also evaluate the factors associated with each effect taken separately
from each other 
\begin{equation}
\eta _{\rm F}^{\rm P}=\frac{F^{\rm P}}{F_{\rm Cas}}\qquad \eta _{%
{\rm F}}^{\rm T}=\frac{F^{\rm T}}{F_{\rm Cas}}  \label{etaF}
\end{equation}
$F^{\rm P}$ is the Casimir force evaluated by accounting for finite
conductivity of the metals but assuming zero temperature and $F^{\rm T}$
is the Casimir force evaluated at temperature $T$ for perfect reflectors. Of
course $\eta _{\rm F}^{\rm P}$ depends on the ratio $\frac{L}{\lambda _{%
{\rm P}}}$ and $\eta _{\rm F}^{\rm T}$ on the ratio $\frac{L}{\lambda _{%
{\rm T}}}$. 

Now the question raised in the previous paragraphs may be stated precisely: 
to which level of accuracy can the complete correction factor $\eta _{\rm F}$ 
be approximated as the product of the factors $\eta _{\rm F}^{\rm P}$ and 
$\eta _{\rm F}^{\rm T}$ ? To answer this question we will evaluate the quantity 
\begin{equation}
\delta _{\rm F}=\frac{\eta _{\rm F}}{\eta _{\rm F}^{\rm P}\eta _{%
{\rm F}}^{\rm T}}-1  \label{deltaF}
\end{equation}
which measures the degree of validity of the approximation where both
effects are evaluated independently from each other. 
We will give an analytical estimation of this deviation
which may thus be taken into account without any difficulty. We will 
also give the same results for the Casimir energy by defining a
factor $\eta _{\rm E}$ measuring the whole correction of Casimir energy
due to conductivity and temperature and then discussing the factors 
$\eta _{\rm E}^{\rm P}$ and $\eta _{\rm E}^{\rm T}$ and the deviation 
$\delta _{\rm E}$ in the same manner as for the force. 

Some additional remarks have to be made at this point. First, recent experiments
are not performed in the plane-plane but in the plane-sphere configuration.
The Casimir force in this geometry is usually estimated from the proximity
theorem \cite{Deriagin68,Blocki77,Mostepanenko85,Bezerra97,Klimchitskaya99}.
Basically this amounts to evaluating the force by adding the contributions
of various distances as if they were independent. In the plane-sphere
geometry the force evaluated in this manner turns out to be given by the
Casimir energy evaluated in the plane-plane configuration for the distance $L
$ being defined as the distance of closest approach in the plane-sphere
geometry. Hence, the factor $\eta _{\rm E}$ evaluated in this paper for
energy can be used to infer the factor for the force measured in the
plane-sphere geometry. Then, surface roughness corrections will not be
considered in the present paper. Finally the dielectric response of the
metallic mirrors will be described by a plasma model. This model is known to
describe correctly the Casimir force in the long distance range which is
relevant for the study of temperature effects. Keeping these remarks in mind,
our results will provide one with an accurate evaluation of the Casimir
force in the whole range of experimentally explored distances.

\section{Casimir force and free energy}

When real mirrors are characterized by frequency dependent reflection
coefficients, the Casimir force is obtained as an integral over frequencies
and wavevectors associated with vacuum and thermal fluctuations \cite
{Jaekel91}. The Casimir force is a sum of two parts corresponding to the 2
field polarizations with the two parts having the same form in terms of the
corresponding reflection coefficients 
\begin{eqnarray}
&&F=\sum_{k=-\infty }^{\infty } \frac {\omega _{\rm T}}{2}\ {\Bbb F}
\left[ k\omega _{\rm T}\right]   \nonumber \\
&&{\Bbb F}\left[ \omega \ge 0\right] =\frac{\hbar A}{2\pi ^{2}}\int_{%
\frac{\omega }{c}}^{+\infty }{\rm d}\kappa \ \kappa ^{2}\ f  \nonumber \\
&&f=\frac{r_{\bot }^{2}\left( i\omega ,i\kappa \right) }{e^{2\kappa
L}-r_{\bot }^{2}\left( i\omega ,i\kappa \right) }+\frac{r_{||}^{2}\left(
i\omega ,i\kappa \right) }{e^{2\kappa L}-r_{||}^{2}\left( i\omega ,i\kappa
\right) }  \nonumber \\
&&{\Bbb F}\left[ -\omega \right] ={\Bbb F}\left[ \omega \right] 
\label{Fexact}
\end{eqnarray}
$r_{\bot }$ (respectively $r_{||}$) denotes the amplitude reflection
coefficient for the orthogonal (respectively parallel) polarization of one
of the two mirrors. The mirrors are here supposed to be identical, otherwise 
$r_{\bot }^{2}$
should be replaced by the product of the two coefficients. $\omega $ is the
frequency and $\kappa $ the wavevector along the longitudinal direction of
the cavity formed by the $2$ mirrors. ${\Bbb F}\left[ \omega \right] $ is 
defined for positive frequencies and extended to negative ones by parity. 

The Casimir force (\ref{Fexact}) may also be rewritten after a Fourier 
transformation, as a consequence of Poisson formula \cite{Schwinger78}
\begin{eqnarray}
F &=&\sum_{m=-\infty }^{\infty } \widetilde{\Bbb F} 
\left( m\lambda _{\rm T} \right)   \nonumber \\
\widetilde{\Bbb F}(x) &=&\int_{0}^{\infty }{\rm d}\omega \ \cos
\left( \frac{\omega x}{c}\right) \ {\Bbb F}\left[ \omega \right] 
\label{Fpoisson}
\end{eqnarray}
The contribution of vacuum fluctuations, that is also the limit
of a null temperature $\left( \omega _{\rm T}\rightarrow 0\right) $ in
(\ref{Fexact}), corresponds to the contribution $m=0$ in (\ref{Fpoisson})
\begin{equation}
F^{\rm P}= \widetilde{\Bbb F} \left( 0 \right) =
\int_{0}^{\infty }{\rm d}\omega \ {\Bbb F}\left[ \omega
\right]   \label{Fplasma}
\end{equation}
Hence, the whole force (\ref{Fpoisson}) is the sum of this vacuum contribution 
$m=0$ and of thermal contributions $m \neq 0$.

We will consider metallic mirrors with the dielectric function $\varepsilon
\left( i\omega \right) $ for imaginary frequencies given by the plasma model 
\begin{equation}
\varepsilon \left( i\omega \right) =1+\frac{\omega _{\rm P}^{2}}{\omega^{2}}
\end{equation}
$\omega _{\rm P}$ is the plasma frequency related to the plasma wavelength 
$\lambda _{\rm P}$ by (\ref{lambdaP}). For the metals used in recent
experiments, the values chosen for the plasma wavelength $\lambda _{\rm P}$
will be 107nm for Al and 136nm for Cu and Au. These values are in agreement
with knowledge from solid state physics \cite{Schulz57,Ehrenreich62} as well
as with the integration of optical data described in detail in \cite
{Lambrecht00}. As already known, the results obtained from the plasma model
departs from the more accurate integration of optical data for small
distances. In this limit however, the thermal corrections do not play a
significant role. In the present paper we will restrict our attention to the
plasma model and discuss the validity of the results obtained in this manner
at the end of the next section.

We will also focus the attention on mirrors with a large optical thickness
for which the reflection coefficients $r_{\bot }\left( i\omega ,i\kappa
\right) $ and $r_{||}\left( i\omega ,i\kappa \right) $ correspond to a
simple vacuum-metal interface. With the plasma model, these coefficients are
read as 
\begin{eqnarray}
r_{\bot } &=&-\frac{\sqrt{\omega _{\rm P}^{2}+c^{2}\kappa ^{2}}-c\kappa }{%
\sqrt{\omega _{\rm P}^{2}+c^{2}\kappa ^{2}}+c\kappa }  \nonumber \\
r_{||} &=&\frac{\sqrt{\omega _{\rm P}^{2}+c^{2}\kappa ^{2}}-c\kappa \left(
1+\frac{\omega _{\rm P}^{2}}{\omega ^{2}}\right) }{\sqrt{\omega _{\rm P}^{2}
+c^{2}\kappa ^{2}}+c\kappa \left( 1+\frac{\omega _{\rm P}^{2}}{\omega
^{2}}\right) }  \label{coeffR}
\end{eqnarray}
For wavevectors $c\kappa $ smaller than $\omega _{\rm P}$, 
mirrors may be considered to be perfectly reflecting. When converted to
the distance domain, this entails that the force approaches
the ideal Casimir expression when evaluated at large distances 
$L \gg \lambda _{\rm P}$.

The Casimir energy will be obtained from the force by integration over the
mirrors relative distance 
\begin{equation}
E=\int_{L}^{\infty }F(x){\rm d}x
\end{equation}
As this procedure is performed at constant temperature, the energy thus
obtained corresponds to the thermodynamical definition of a free energy.
For simplicity we will often use the denomination of an energy. We will
define a factor $\eta _{\rm E}$ measuring the whole correction of energy
due to conductivity and temperature effects with respect to the ideal
Casimir energy 
\begin{eqnarray}
\eta _{\rm E} &=&\frac{E}{E_{\rm Cas}}  \nonumber \\
E_{\rm Cas} &=& \frac{\hbar c A \pi ^{2}}{720L^{3}}
\label{defEtaE}
\end{eqnarray}
The positive value of the energy here means that the Casimir energy is a binding
energy while the positive value of the force is associated with an
attractive character. We will then define $2$ factors $\eta _{\rm E}
^{\rm P}$ and $\eta _{\rm E}^{\rm T}$ associated with each effect taken
separately from each other, as in (\ref{etaF}). As already done for the
force correction factors in (\ref{deltaF}), we will finally evaluate the
quantity $\delta _{\rm E}$ which characterizes
the degree of validity of the approximation where both effects are evaluated
independently from each other. As mentioned in the Introduction, the results
obtained for energy allows one to deal with the Casimir force in the
plane-sphere geometry when trusting the proximity force theorem.

\section{Numerical evaluations}

In the following we present the numerical evaluation of the correction
factors of the Casimir force and energy using equations written in the
former section.

The force correction factor 
was evaluated for the experimentally relevant distance range of 
0.1-10$\mu$m with the help of equation (\ref{Fpoisson}), supposing explicitly 
a plasma model for the dielectric function, and the result was normalized by 
the ideal Casimir force. A double integration over frequencies and wavevectors 
had to be performed. Due to the cosine dependence in (\ref{Fpoisson}), the 
integrand turned out to be a highly oscillating function. Hence, the integration 
required care although it was performed with standard numerical routines. 
The energy correction factor was then calculated by numerically integrating 
the force and normalizing by the ideal Casimir energy (see equation 
(\ref{defEtaE})). Integration was restricted to a finite interval, the upper 
limit exceeding at least by a factor of $10^4$ the distance at which the 
energy value was calculated. Extending the integration range by a factor of 
100 changed the numerical result by less than $10^{-7}$.

\begin{figure}[tbh]
\centerline{\psfig{figure=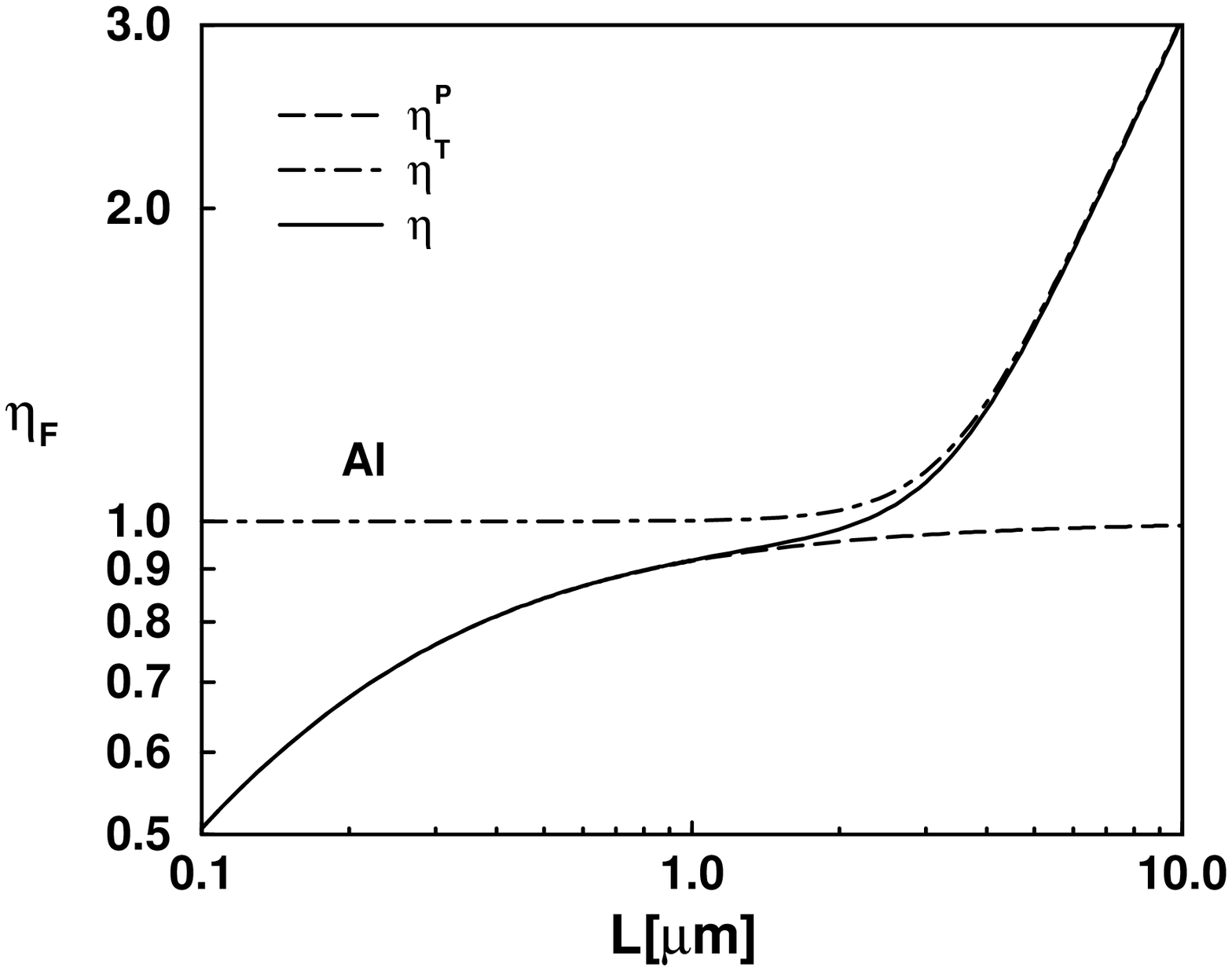,width=8cm}}
\centerline{\psfig{figure=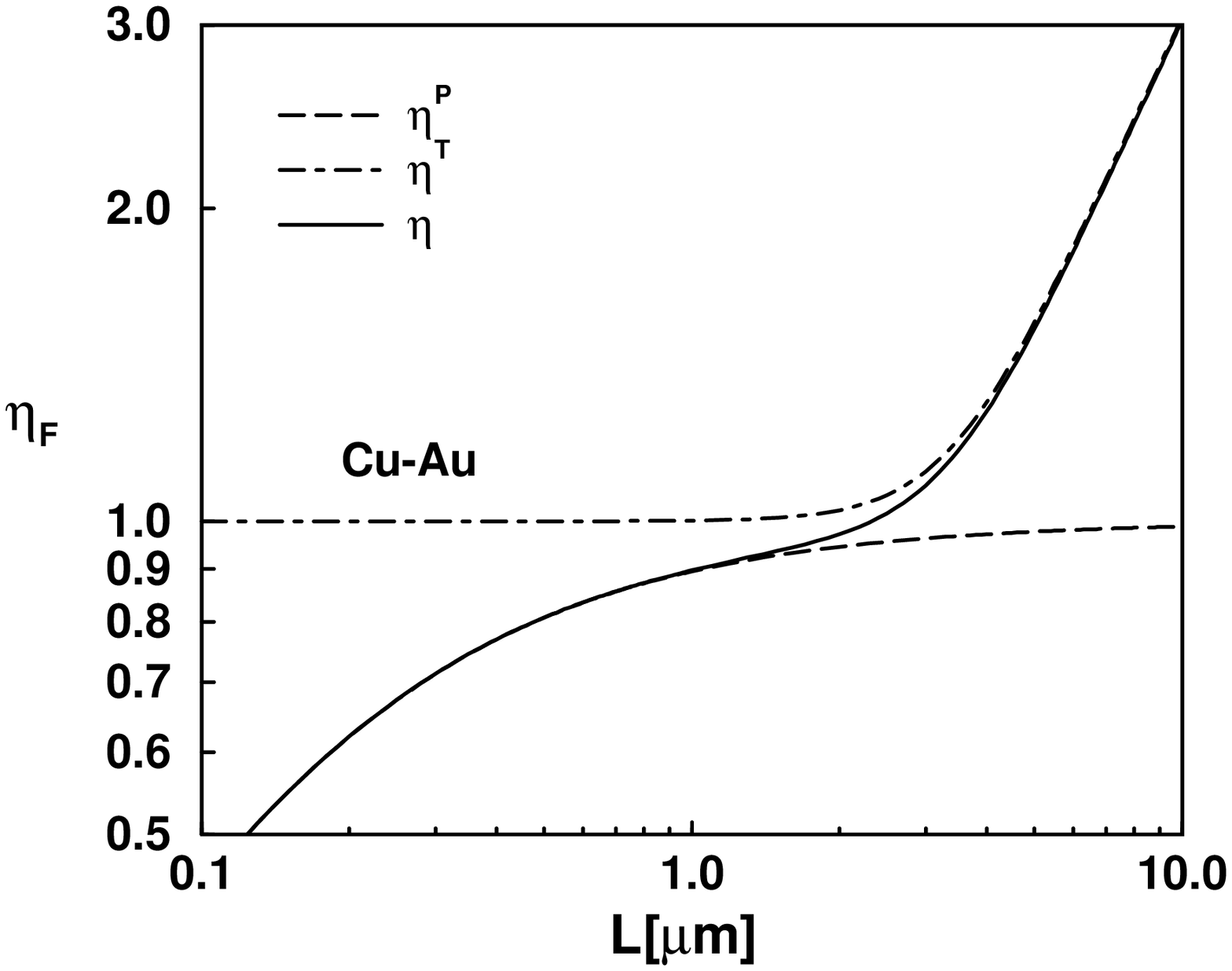,width=8cm}}
\caption{Force correction factor for Al (upper figure) and Cu and Au (lower
graph) as function of the mirrors distance at $T=300K$.}
\label{fig1}
\end{figure}

The results of the numerical evaluation of $\eta _{\rm F}$ are shown 
as the solid lines in figures \ref{fig1} for Al and for Cu-Au assuming a
temperature of $T=300K$. They are compared with the force reduction factor 
$\eta _{\rm F}^{\rm P}$ due to finite conductivity (dashed lines) and the
force enhancement factor $\eta _{\rm F}^{\rm T}$ calculated for perfect
mirrors at 300K (dashed-dotted lines).

\begin{figure}[tbh]
\centerline{\psfig{figure=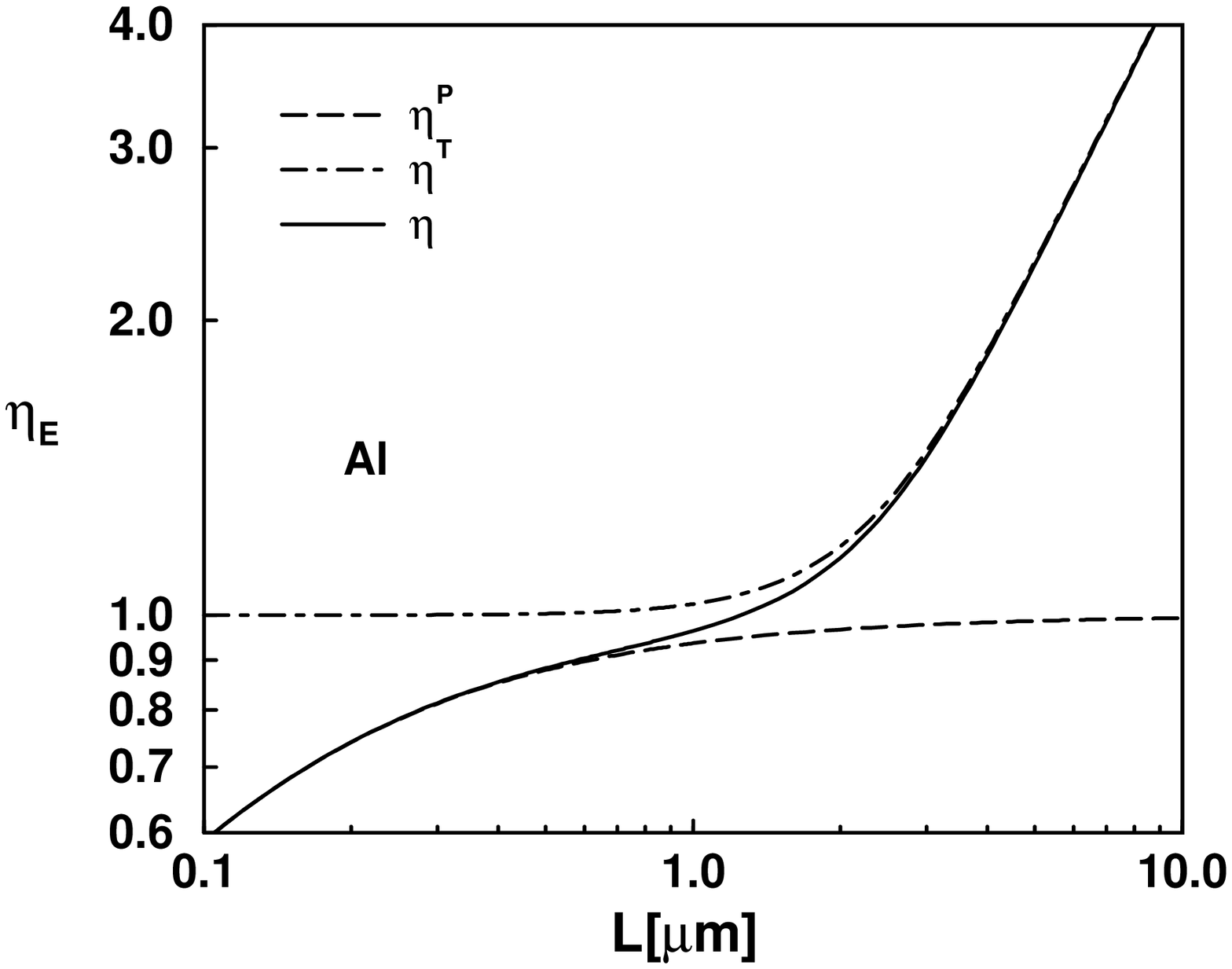,width=8cm}}
\centerline{\psfig{figure=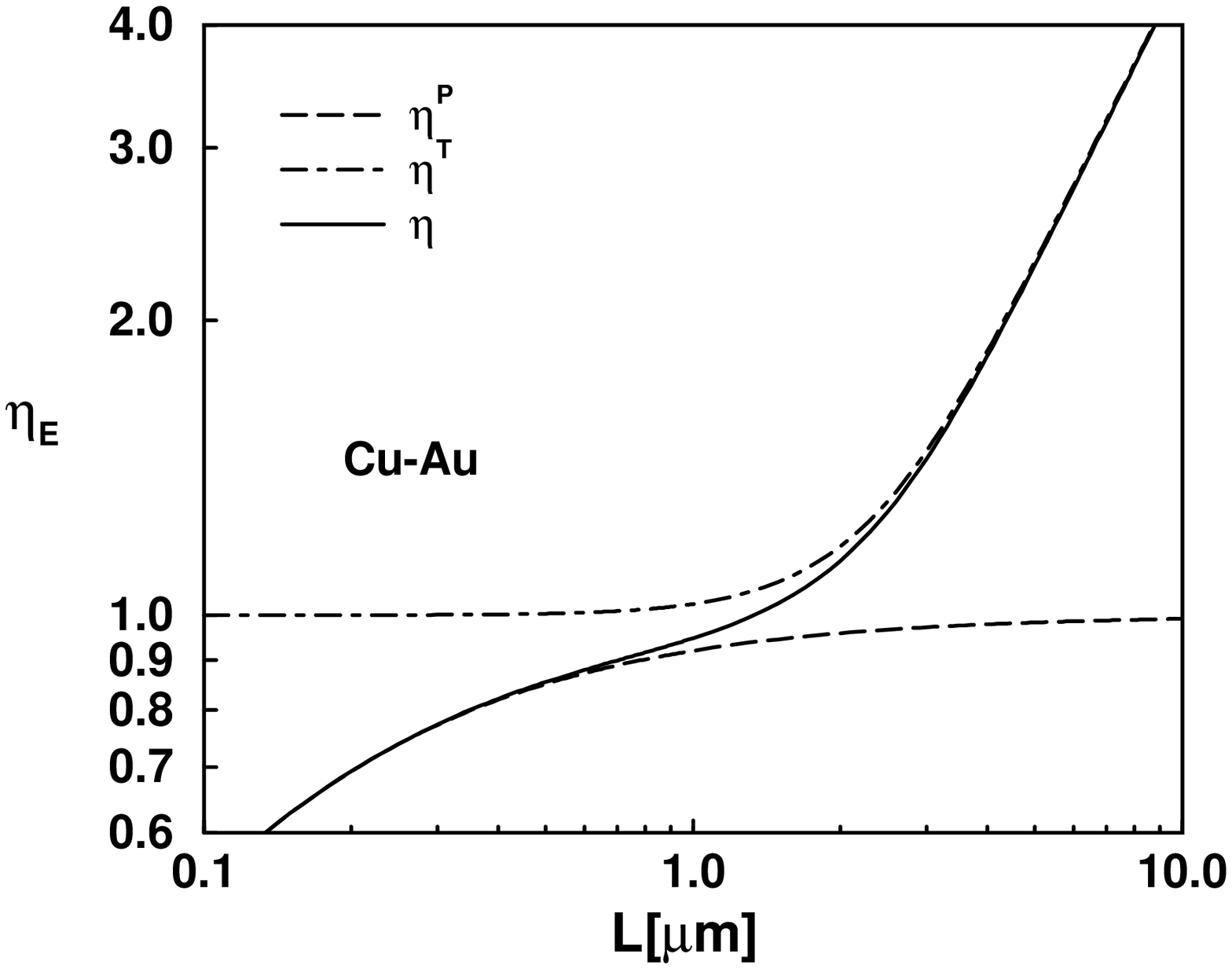,width=8cm}}
\caption{Energy correction factor for Al (upper figure) and Cu and Au (lower
graph) as function of the mirrors distance at $T=300K$.}
\label{fig2}
\end{figure}

Figure \ref{fig2} shows similar results for the factor $\eta _{\rm E}$
obtained through numerical evaluation of the Casimir free energy. The shape
of the graphs is similar to the ones of the force. However, while finite
conductivity corrections are more important for the force, thermal effects
have a larger influence on energy. 

\begin{figure}[tbh]
\centerline{\psfig{figure=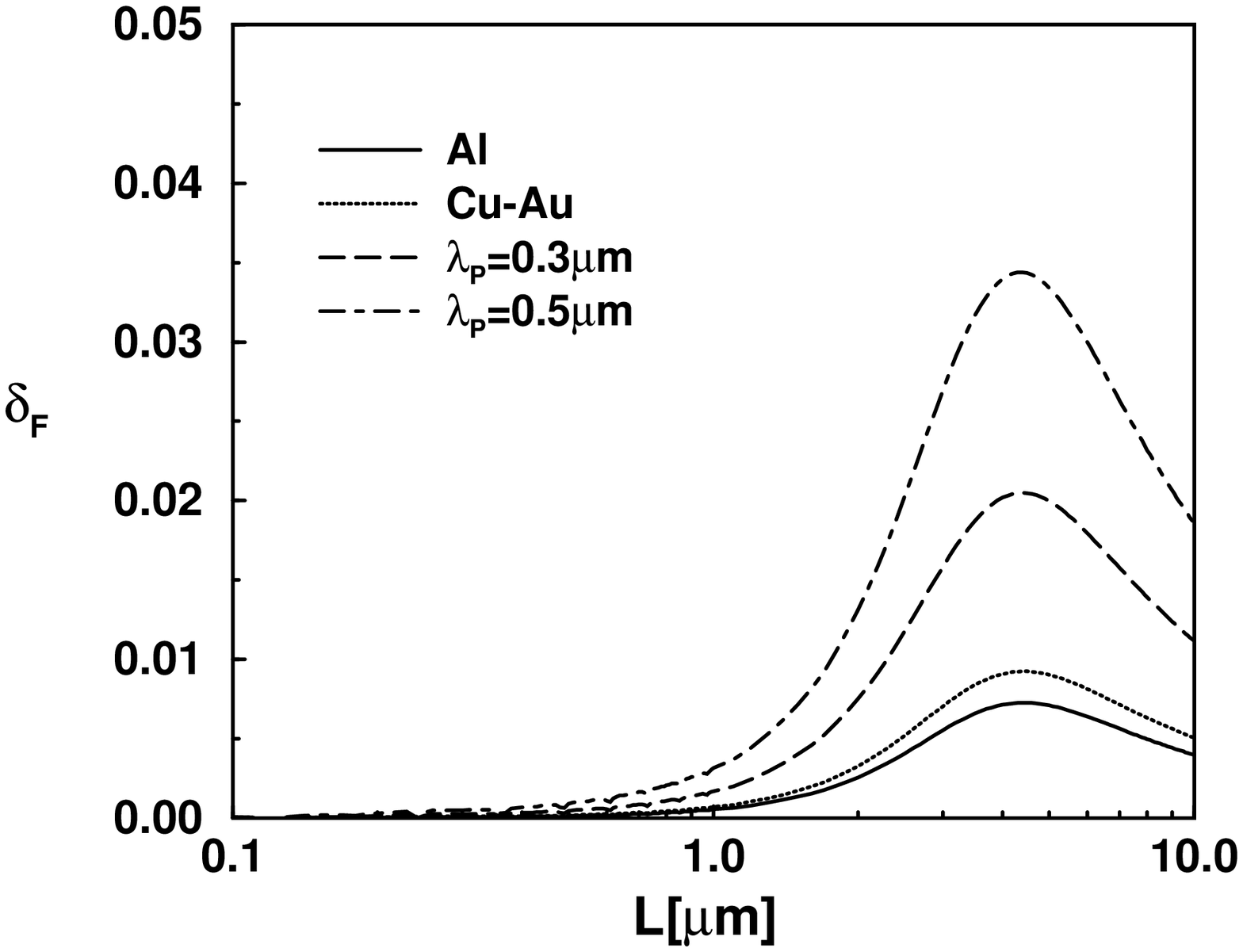,width=8cm}}
\centerline{\psfig{figure=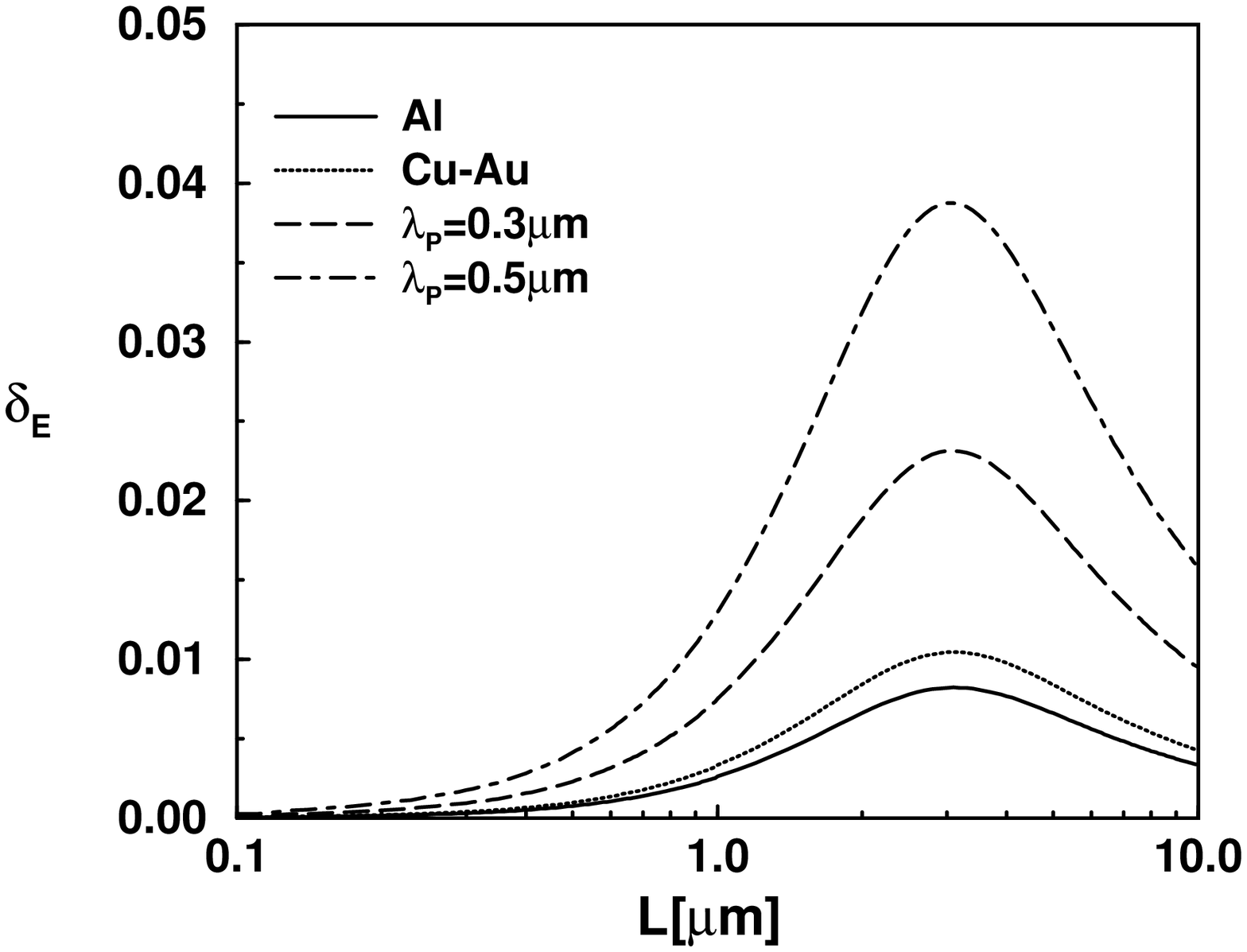,width=8cm}}
\caption{$\delta _{\rm F}$ (upper graph) and $\delta _{\rm E}$
(lower graph) as a function of the mirrors distance. The results are given
for the three metals Al, Cu-Au and two larger plasma wavelengths.}
\label{fig3}
\end{figure}

For the force as well as for the energy, temperature corrections are
negligible in the short distance limit while conductivity corrections may be
ignored at large distances. The whole correction factor $\eta $ behaves
roughly as the product $\eta ^{\rm P}\eta ^{\rm T}$ of the 2 correction
factors evaluated separately. However, both correction factors are
appreciable in the distance range $1-4\mu $m in between the two
limiting cases. Since this range is important for the comparison between
experiments and theory, it is necessary to discuss in a more precise manner
how good is the often used approximation which identifies $\eta $ to the
product $\eta ^{\rm P}\eta ^{\rm T}$. In order to assess the quality of
this approximation, we have plotted in figure \ref{fig3} the quantities $%
\delta _{\rm F}$ and $\delta _{\rm E}$ as a function of the distance for
Al, Cu-Au and two additional plasma wavelengths. 
A value of $\delta =0$ would signify that the approximation gives an
exact estimation of the whole correction. An important outcome of our
calculation is that the errors $\delta _{\rm F}$ and $\delta _{\rm E}$
are of the order of 1\% for Al and Cu-Au at a temperature of $300K$. 

For estimations at the 5\% level, the separate calculation of $\eta ^{\rm P}$ 
and $\eta ^{\rm T}$ and the evaluation of $\eta $ as the product $\eta
^{\rm P}\eta ^{\rm T}$ can therefore be used. However, if a 1\% level
or a better accuracy is aimed at, this approximation is not sufficient. It
should be noticed furthermore that the error increases when the temperature
or the plasma wavelength are increased. It becomes of the order of 4\% for a
plasma wavelength of 0.5 $\mu $m at 300K. The sign obtained for $\delta $
means that the approximation gives too small values of force and energy.  

We want now to emphasize a few points. In order to make the discussion
precise, we give numerical values of the correction factors for $2$ 
experimentally relevant distances, namely $0.5 \mu$m and $3 \mu$m.
The first distance corresponds to the smallest distance for which
the plasma model gives results in correct agreement with the integration 
of optical data \cite{Lambrecht00}. For this distance, the thermal 
corrections do not play a significant role 
($\eta _{\rm F}^{\rm T} = 1.000$; $\eta _{\rm E}^{\rm T} = 1.004$). 
\begin{equation}
\begin{array}{cccc}
\quad L=0.5 \mu {\rm m}\quad  & \qquad &  &  \\
& \quad {\rm Al}\quad  & \quad {\rm Cu-Au}\quad  &  \\[0.5mm] 
\eta _{\rm F}^{\rm P}  & 0.843 & 0.808 &  \\[0.5mm]
\eta _{\rm F}  & 0.843 & 0.808 &  \\ [1mm]
\eta _{\rm E}^{\rm P}  & 0.879 & 0.851 &  \\[0.5mm] 
\eta _{\rm E}  & 0.883 & 0.855 &
\end{array}
\end{equation}
At shorter distances the results obtained with the plasma model depart 
from the values calculated from the integration of optical data by more 
than 1\%. Hence, the values of $\eta _{\rm F}^{\rm P}$ and 
$\eta _{\rm E}^{\rm P}$ used for distances smaller than 0.5$\mu $m 
have to take into account the more accurate dielectric function obtained 
through an integration of optical data \cite{Lambrecht00}. 
In this short distance range however, the whole correction factors 
$\eta _{\rm F}$ and $\eta _{\rm E}$ may be obtained as the products 
$\eta _{\rm F}^{\rm P} \eta _{\rm F}^{\rm T}$
and $\eta _{\rm E}^{\rm P} \eta _{\rm E}^{\rm T}$.

In the long distance range in contrast, the temperature correction becomes
predominant. The conductivity correction has still to be accounted for but
it may be calculated by using the plasma model. This is illustrated
by the correction factors obtained for a distance of 3$\mu $m  
($\eta _{\rm F}^{\rm T} = 1.117$; $\eta _{\rm E}^{\rm T} = 1.470$). 
\begin{equation}
\begin{array}{cccc}
\quad L= 3 \mu {\rm m}\quad  & \qquad &  &  \\
& \quad {\rm Al}\quad  & \quad {\rm Cu-Au}\quad  &  \\[0.5mm]
\eta _{\rm F}^{\rm P}  & 0.971 & 0.963 &  \\[0.5mm]
\eta _{\rm F}^{\rm P} \eta _{\rm F}^{\rm T}  & 1.084 & 1.076 &  \\[0.5mm] 
\eta _{\rm F} & 1.090 & 1.083 &  \\[1mm]
\eta _{\rm E}^{\rm P}  & 0.978 & 0.972 &  \\[0.5mm] 
\eta _{\rm E}^{\rm P} \eta _{\rm E}^{\rm T}  & 1.437 & 1.429 &  \\[0.5mm] 
\eta _{\rm E} & 1.449 & 1.444 &   
\end{array}
\end{equation}
For this distance, all corrections have to be taken into account. 
The metals cannot be considered as perfect reflectors yet, the temperature
corrections are significant and the deviation between the exact 
correction and the mere product has to be included if a high accuracy 
is aimed at. This is especially true in the case of Casimir energy.

\section{Scaling laws for the deviations}

An inspection of figure \ref{fig3} shows that the curves corresponding to
different plasma wavelengths $\lambda _{\rm P}$ have similar shapes with a
maximum which is practically attained for the same distance between the
mirrors. The amplitude of the deviations, which is larger for the energy than for 
the force, is found to vary linearly as a function of the plasma wavelength 
$\lambda _{\rm P}$. 

\begin{figure}[tbh]
\centerline{\psfig{figure=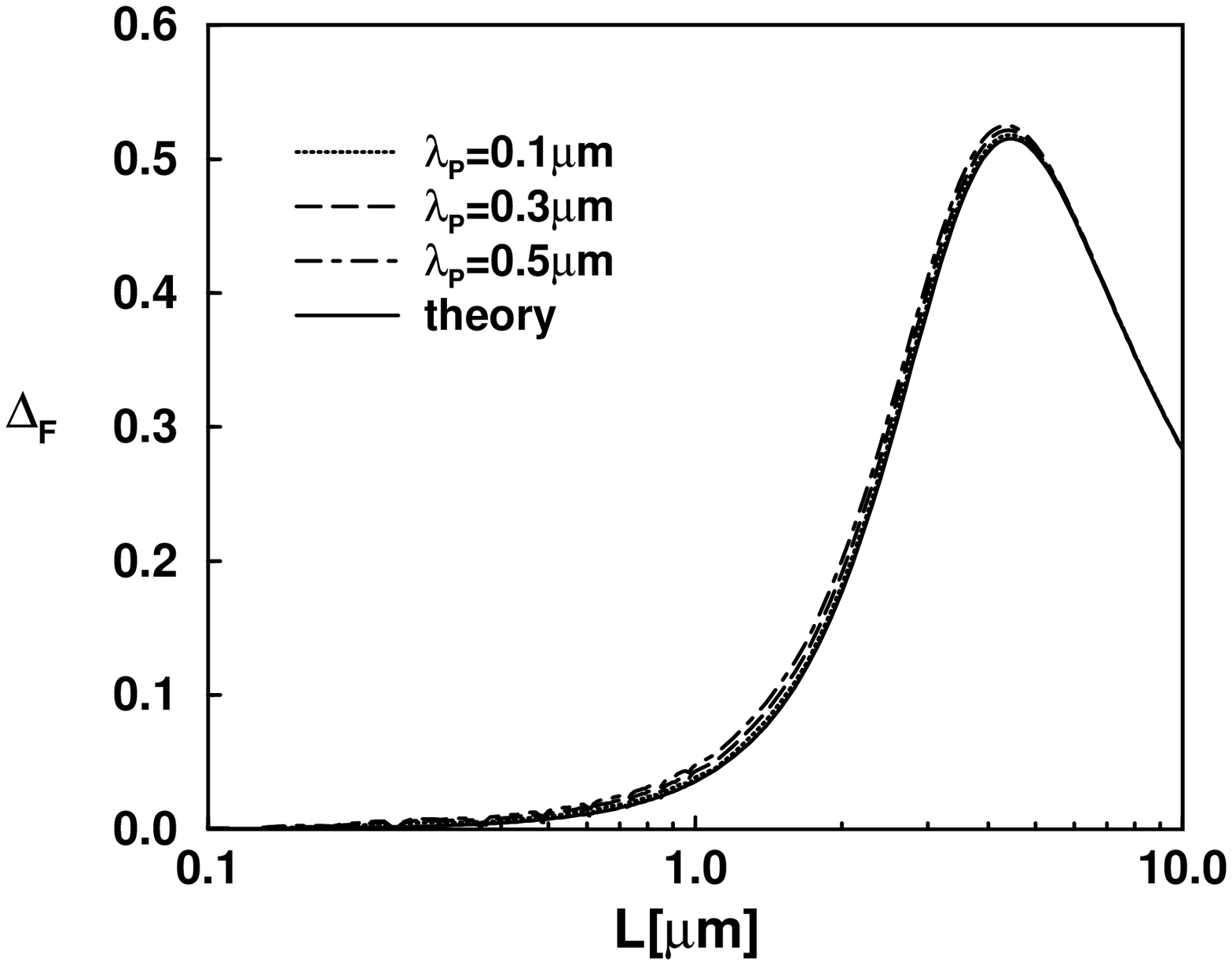,width=8cm}}
\centerline{\psfig{figure=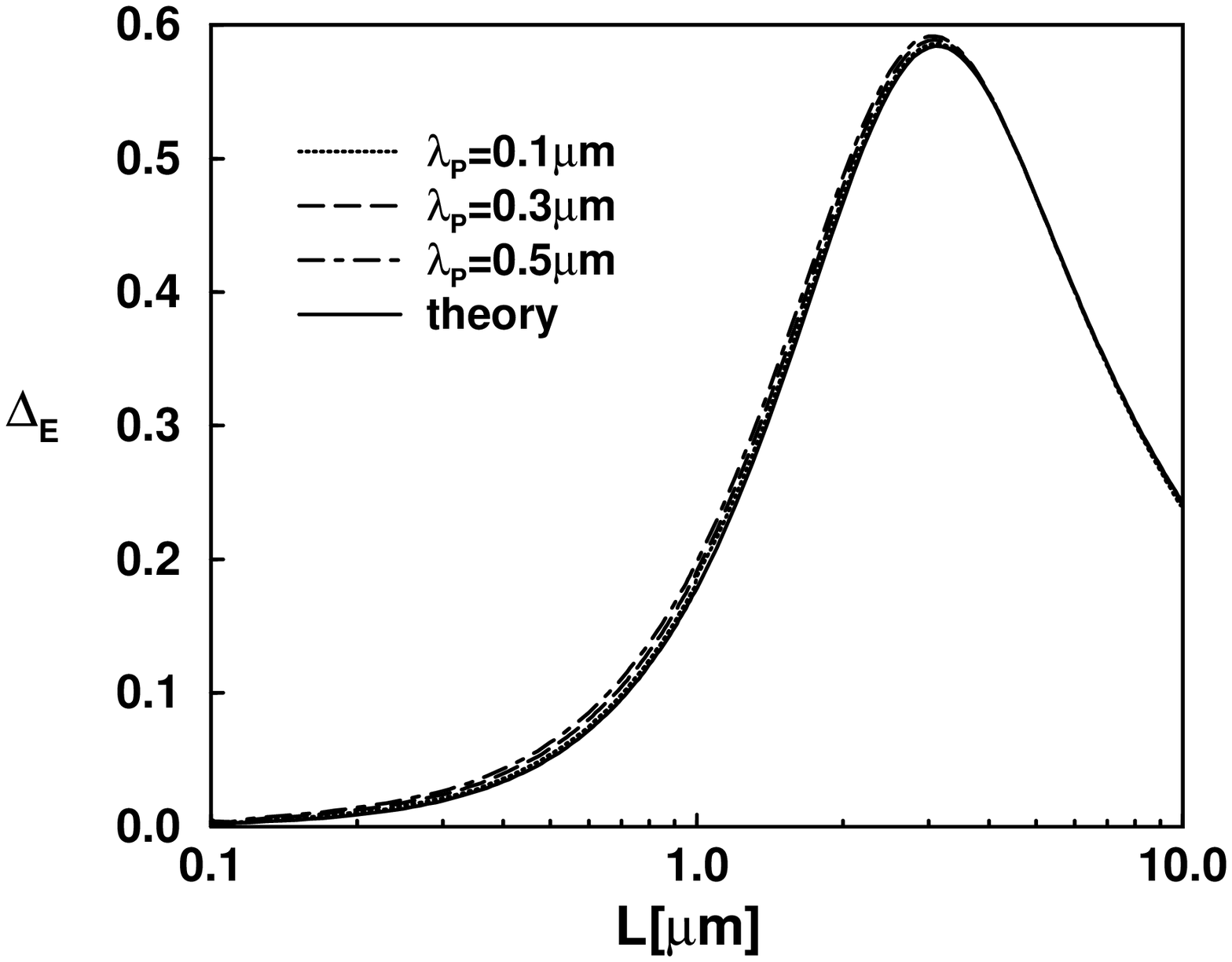,width=8cm}}
\caption{The deviations are represented for the force (upper
graph) and the free energy (lower graph) after the rescaling described
by equation (\ref{Scaling}). Different plasma wavelengths lead to nearly
identical functions, drawn as dotted, dashed and dotted-dashed lines. These
functions are hardly distinguishable from the solid lines which represent
the analytical expressions derived in the next section.}
\label{fig4}
\end{figure}

This scaling property is confirmed by figure \ref{fig4} where we have drawn
the deviations after an appropriate rescaling
\begin{equation}
\Delta =\frac{\lambda _{\rm T}}{\lambda _{\rm P}}\delta  \label{Scaling}
\end{equation}
The curves obtained for $\Delta _{\rm F}$ and $\Delta _{\rm E}$ for
different plasma wavelengths at temperature $T=300K$ are nearly perfectly
identical to each other. These curves correspond to values of the plasma 
wavelength small compared to the thermal wavelength and the scaling law
would not be obeyed so well otherwise. 

In other words, the deviations $\delta _{\rm F}$ and $\delta _{\rm E}$
are proportional to the factor $\frac{\lambda _{\rm P}}{\lambda _{\rm T}}
$ on one hand, and to the functions $\Delta _{\rm F}$ and $\Delta _{\rm E}$ 
on the other hand. The latter functions, which no longer depend on 
$\lambda _{\rm P}$, provide a simple method for reaching a good accuracy in
the theoretical estimation of the whole correction factor  
\begin{equation}
\eta =\eta ^{\rm P}\eta ^{\rm T}\left( 1+\frac{\lambda _{\rm P}}{%
\lambda _{\rm T}}\Delta \right) 
\end{equation}
This method is less direct than the complete numerical integration of the
forces which has been performed for obtaining the curves presented in the
previous section. But it requires easier computations while nevertheless 
giving accurate estimations of the correction factors. Typically, the 
deviation $\delta$ with a magnitude of the order of the \% may be estimated 
with a much better precision through the mere inspection of figure \ref{fig4}. 
Alternatively, one may use the analytical expression of the functions $\Delta$ 
presented in the next section and drawn as the solid lines on figure \ref{fig4}. 

\section{Analytical expressions of the deviations}

The results of numerical integrations presented in the foregoing section
have shown that the deviations $\delta _{\rm F}$ and $\delta _{\rm E}$
are proportional to the plasma wavelength $\lambda _{\rm P}$,
for plasma wavelengths small compared to the thermal wavelength. 
In this final section, we explain this scaling law by
using a partial analytical integration of the whole correction factors.

To this aim, we write the force correction factor by dividing 
(\ref{Fpoisson}) by the value of the ideal Casimir force 
\begin{equation}
\eta _{\rm F}=\eta _{\rm F}^{\rm P}+\left( \eta _{\rm F}^{\rm T}
-1\right) +\Delta \eta _{\rm F}  \label{etaSum}
\end{equation}
The first term in (\ref{etaSum}) corresponds to the vacuum contribution 
(\ref{Fplasma}) 
\begin{equation}
\eta _{\rm F}^{\rm P}=\frac{120L^{4}}{\pi ^{4}}
\int_{0}^{\infty }{\rm d}\kappa \ \kappa ^{3}
\int_{0}^{1}{\rm d}y\ f  \label{etaFplasma}
\end{equation}
with $f$ still given by (\ref{Fexact}).
A dimensionless frequency $y=\frac{\omega }{c\kappa }$ measured with 
respect to the wavevector has been introduced. Note also that the 
wavevector $\kappa$ is involved through the dimensionless quantity 
$\kappa L$, except in the expressions of reflection coefficients. In
(\ref{etaFplasma}), the integration over $y$ may be performed analytically 
(see the appendix \ref{appVacuum}). At long distances, $\eta _{\rm F}^{\rm P}$
tends to the limit of perfect reflection with a known correction \cite{Schwinger78} 
\begin{equation}
L\gg \lambda _{\rm P}\quad \rightarrow \quad \eta _{\rm F}^{\rm P}=1-%
\frac{8}{3\pi }\frac{\lambda _{P}}{L}+\ldots   \label{etaPlong}
\end{equation}
This expansion has been the subject of a number of papers and it has been used
to propose interpolation formulas \cite{Bezerra97,Klimchitskaya99}. However
such a series expansion can hardly reproduce the behavior at small
distances where $\eta _{\rm F}^{\rm P}$ varies as $\frac{L}{\lambda_{P}}$, 
which just means that the conductivity effect is not a small perturbation 
at short distances (see the appendix \ref{appVacuum}). 

Coming back to the whole expression (\ref{etaSum}) of the force correction
factor, it remains to discuss the thermal contributions, that is the second
and third terms. These two terms come from the contributions 
$m\neq 0$ to (\ref{Fpoisson}). The opposite values of $m$ give equal 
contributions and they have been gathered. The thermal contributions have
been split in two parts, the second and third terms in (\ref{etaSum}), 
which correspond respectively to the limit of perfect mirrors on one hand 
\begin{eqnarray}
\eta _{\rm F}^{\rm T}-1 &=& \frac{240L^{4}}{\pi ^{4}} 
\sum_{m=1}^{\infty }
\int_{0}^{\infty }{\rm d}\kappa \ \kappa ^{3}
\int_{0}^{1}{\rm d}y\ \cos \left( my\kappa \lambda _{\rm T}
\right)  f_1   \nonumber \\
f_1 &=& \frac{2}{e^{2\kappa L}-1}  \label{etaFT}
\end{eqnarray}
and the remainder on the other hand 
\begin{eqnarray}
\Delta \eta _{\rm F} &=& \frac{240L^{4}}{\pi ^{4}} \sum_{m=1}^{\infty }
\int_{\frac{\omega }{c}}^{\infty }{\rm d}\kappa \ \kappa ^{3}
\int_{0}^{1}{\rm d}y\ \cos \left( my\kappa \lambda _{\rm T}\right)
\Delta f  \nonumber \\
\Delta f &=&f-f_1   \nonumber \\
&=&-\frac{e^{2\kappa L}}{e^{2\kappa L}-1}\left( \frac{1-r_{\bot }^{2}}{%
e^{2\kappa L}-r_{\bot }^{2}}+\frac{1-r_{||}^{2}}{e^{2\kappa L}-r_{||}^{2}}%
\right)   \label{etaFD}
\end{eqnarray}
The contribution (\ref{etaFT}) has been denoted $\left( \eta _{\rm F}^{%
{\rm T}}-1\right) $ with $\eta _{\rm F}^{\rm T}$ the correction
factor obtained for perfect mirrors at a non zero temperature. For this term
the integration over $y$ is trivial and the integration over $\kappa $ may
be performed analytically, leading to the known expression 
\cite{Mehra67,Brown69,Schwinger78} 
\begin{eqnarray}
\eta _{\rm F}^{\rm T}-1 &=&\frac{480L^{4}}{\pi ^{4}}\
\sum_{m=1}^{\infty }\ \int_{0}^{\infty }{\rm d}\kappa \ \frac{\kappa
^{2}}{e^{2\kappa L}-1}\ \frac{\sin \left( m\kappa \lambda _{\rm T}\right) 
}{m\lambda _{\rm T}}  \nonumber \\
&=&30\sum_{m=1}^{\infty }\left( \frac{1}{\left( \alpha m\right) ^{4}}-\frac{%
{\rm \cosh }\left( \alpha m\right) }{\alpha m\ {\rm \sinh }^{3}\left( \alpha
m\right) }\right)   \nonumber \\
\alpha  &=&\frac{\pi \lambda _{\rm T}}{2L}
\end{eqnarray}

To obtain the overall correction factor (\ref{etaSum}) it now remains to
evaluate the last expression (\ref{etaFD}). This can be done numerically,
thus leading to the same results as in the previous section since no
approximation has been performed up to now. But the results of the previous
section suggest that we may obtain an accurate estimation of this term through 
an expansion in powers of $\lambda _{\rm P}$. The plasma wavelength 
$\lambda _{\rm P}$ is indeed much smaller than the thermal wavelength 
$\lambda _{\rm T}$ in all experimental situations studied up to now. 
Also, the deviation studied in the foregoing section is appreciable
only for distances $L$ much larger than $\lambda _{\rm P}$. 
Hence an accurate description of the deviation factor 
should be obtained by evaluating $\Delta \eta _{\rm F}$ at the first order 
in $\lambda _{\rm P}$. 

This first order term is easily deduced from (\ref{coeffR},\ref{etaFD}) 
\begin{eqnarray}
\Delta f &\simeq &-\frac{e^{2\kappa L}}{\left( e^{2\kappa L}-1\right) ^{2}}%
\left( 1-r_{\bot }^{2}+1-r_{||}^{2}\right)   \nonumber \\
&\simeq &-\frac{e^{2\kappa L}}{\left( e^{2\kappa L}-1\right) ^{2}}\frac{%
2\kappa \lambda _{\rm P}}{\pi }\left( 1+y^{2}\right) 
\end{eqnarray}
It is proportional to $\lambda _{\rm P}$ and to a function $\phi _{\rm F}$ which
does no longer depend on $\lambda _{\rm P}$ 
\begin{eqnarray}
\Delta \eta _{\rm F} &\simeq &\frac{\lambda _{\rm P}}{L}\phi _{\rm F}  \nonumber \\
\phi _{\rm F} &=&\frac{15}{\pi}\sum_{m=1}^{\infty }\left( \frac{{\rm \cosh }\left(
\alpha m\right) }{\left( \alpha m\right) ^{3}{\rm \sinh }\left( \alpha
m\right) }+\frac{1}{\left( \alpha m\right) ^{2}{\rm \sinh }^{2}\left( \alpha
m\right) }\right.   \nonumber \\
&&+\left. \frac{4{\rm \cosh }\left( \alpha m\right) }{\alpha m\ {\rm \sinh }%
^{3}\left( \alpha m\right) }-\frac{2+4{\rm \cosh }^{2}\left( \alpha m\right) 
}{{\rm \sinh }^{4}\left( \alpha m\right) }\right) 
\end{eqnarray}
Collecting the results obtained up to now, we get an estimation of the
force correction factor $\eta _{\rm F}$ valid in the long distance range 
$L\gg \lambda _{\rm P}$ 
\begin{eqnarray}
\eta _{\rm F} &=&\eta _{\rm F}^{\rm P}\eta _{\rm F}^{\rm T}
+\left( 1-\eta _{\rm F}^{\rm P}\right) \left( \eta _{\rm F}^{\rm T}
-1\right) +\Delta \eta _{\rm F}  \nonumber \\
&\simeq &\eta _{\rm F}^{\rm P}\eta _{\rm F}^{\rm T}+
\frac{8}{3 \pi} \frac{\lambda _{P}}{L} \left( \eta _{\rm F}^{\rm T}
-1\right) + \frac{\lambda _{P}}{L} \phi _{\rm F}  
\end{eqnarray}
Coming back to the notations of the previous section, this result is equivalent 
to the following expression for the function $\Delta _{\rm F} $
\begin{eqnarray}
\Delta _{\rm F} &=& \frac{8}{3 \pi} \frac{\lambda _{T}}{L} 
\frac{\eta _{\rm F}^{\rm T}-1}{\eta _{\rm F}^{\rm T}}
+ \frac{\lambda _{T}}{L} \frac{\phi _{\rm F}}{\eta _{\rm F}^{\rm T}}
\end{eqnarray}
This function is plotted as the solid line on figure \ref{fig4} and it is
found to fit well the results of the complete numerical integration 
presented in the previous section.

Similar manipulations can be done for evaluating correction factors for
the Casimir free energy. We give below the main results, that is the 
thermal correction factor evaluated for perfect mirrors 
\begin{eqnarray}
\eta _{\rm E}^{\rm T}-1 
&=&45 \sum_{m=1}^{\infty }\left( - \frac{2}{\left( \alpha m \right) ^{4}}
+\frac{1} {\left( \alpha m \right) ^{3} {\rm \tanh } \left( \alpha m \right) } 
\right. \nonumber \\
&&+\left. \frac{1}{\left( \alpha m \right)^2 {\rm \sinh }^{2}\left( \alpha m\right)}
\right)   
\end{eqnarray}
and the first order correction 
\begin{eqnarray}
\Delta \eta _{\rm E} &\simeq &\frac{\lambda _{\rm P}}{L}\phi _{\rm E}  \nonumber \\
\phi _{\rm E} &=&\frac{45}{\pi} \sum_{m=1}^{\infty }
\left( - \frac{4} {\left( \alpha m \right) ^{4}} + \frac{1}{\left( \alpha m\right) ^{3}
{\rm \tanh }\left( \alpha m\right) } \right.   \nonumber \\
&&+\left. \frac{1}{\left( \alpha m \right)^2 {\rm \sinh }^{2}\left( \alpha m\right)}
+ \frac{2 {\rm \cosh }\left( \alpha m\right)}
{\alpha m \ {\rm \sinh }^{3}\left( \alpha m \right) } \right) 
\end{eqnarray}
Since the long distance expansion of $\eta_{\rm E}^{\rm P}$ up to first order in the
plasma wavelength is given by
\begin{equation}
L\gg \lambda _{\rm P}\quad \rightarrow \quad 
\eta_{\rm E}^{\rm P} = 1 - \frac{2}{\pi}\frac{\lambda_{\rm P}}{L} + \ldots
\end{equation}
we deduce the function $\Delta _{\rm E}$ 
\begin{eqnarray}
\Delta _{\rm E} &=& \frac{2}{\pi} \frac{\lambda _{T}}{L} 
\frac{\eta _{\rm E}^{\rm T}-1}{\eta _{\rm E}^{\rm T}}
+ \frac{\lambda _{T}}{L} \frac{\phi _{\rm E}}{\eta _{\rm E}^{\rm T}}
\end{eqnarray}
This function is plotted as the solid line on the second graph of figure \ref{fig4} 
and also found to fit well the results of the numerical integration.

\section{Summary}

In the present paper, we have given an accurate evaluation of the Casimir
force and Casimir free energy between $2$ plane metallic mirrors, taking
into account conductivity and temperature corrections at the same
time. The whole corrections with respect to the ideal Casimir formulas,
corresponding to perfect mirrors in vacuum, have been characterized by
factors $\eta _{\rm F}$ for the force and $\eta _{\rm E}$ for the
energy. These factors have been computed through a numerical evaluation of
the integral formulas. They have also been given a simplified form as a
product of $3$ terms, namely the reduction factor associated with
conductivity at null temperature, the increase factor associated with
temperature for perfect mirrors, and a further deviation factor measuring a kind of
interplay between the two effects. This last factor turns out to lie in the
1\% range for metals used in the recent experiments performed at ambient
temperature. Hence the conductivity and temperature corrections
may be treated independently from each other and simply multiplied for
theoretical estimations above this accuracy level. 

However, when accurate comparisons between experimental and theoretical
values of the Casimir force are aimed at, the deviation factor 
has to be taken into account in theoretical estimations. 
The deviation factor is appreciable for distances greater than the plasma
wavelength $\lambda _{\rm P}$ but smaller or of the order of the thermal
wavelength $\lambda _{\rm T}$. We have used this property to derive a
scaling law of the deviation factor. This law allows one to obtain a
simple but accurate estimation of the Casimir force and free energy through 
a mere inspection of figure \ref{fig4}. Alternatively one can use analytical
expressions which have been obtained through a first order expansion in 
$\lambda _{\rm P}$ of the thermal contributions to Casimir
forces and fit well the results of complete numerical integration. 

We have
represented the optical properties of metals by the plasma model. This model
does not lead to reliable estimations of the forces at small distances but
this deficiency may be corrected by using the real dielectric function of
the metals. This does not affect the discussion of the present paper, except
for the fact that the pure conductivity effect has to be computed through an
integration of optical data for distances smaller than 0.5$\mu $m. Finally 
surface roughness corrections, which have 
not been considered in the present paper, are expected to play a 
significant role in theory-experiment comparisons in the short distance range.

\vspace{5mm}
\noindent {\bf Acknowledgements} 

We wish to thank Ephraim Fischbach,
Marc-Thierry Jaekel, David Koltick, Paulo Americo Ma\"{\i}a Neto and Roberto Onofrio 
for stimulating discussions.

\appendix 

\section{The vacuum contribution}

\label{appVacuum}In the present appendix, we give further analytical 
expressions for the correction factor $\eta _{\rm F}^{\rm P}$ due to 
conductivity, calculated with the plasma model for a null temperature.

Introducing the notations  
\begin{eqnarray}
\rho  &=&\frac{\sqrt{\omega _{\rm P}^{2}+c^{2}\kappa ^{2}}-c\kappa }{\sqrt{%
\omega _{\rm P}^{2}+c^{2}\kappa ^{2}}+c\kappa } \qquad
y = \frac{\omega}{c \kappa}    \nonumber
\end{eqnarray}
we rewrite the reflection coefficients (\ref{coeffR}) 
\begin{eqnarray}
r_{\bot } &=&-\rho   \qquad
r_{||} =\rho \frac{y^{2}\left( 1-\rho \right) -2}{y^{2}\left( 1-\rho
\right) +2\rho }    \nonumber
\end{eqnarray}
In this case one integration may be performed analytically in (\ref{etaFplasma})
\begin{eqnarray}
&&\eta _{\rm F}^{\rm P}=\frac{120 L^4}{\pi ^{4}}\int_{0}^{\infty }%
{\rm d}\kappa \kappa^{3}\frac{2\rho ^{2}+\rho e^{\kappa L}g}
{e^{2\kappa L}-\rho ^{2}}  \nonumber \\
g &=&\frac{1+a_{-}^{2}}{a_{-}}\arctan \frac{1}{a_{-}}-\frac{1+a_{+}^{2}}{%
a_{+}}\arctan \frac{1}{a_{+}}  \nonumber \\
a_{\pm} &=&\sqrt{\frac{e^{\kappa L}\pm \rho}{e^{\kappa L} \mp \rho}
\frac{1+\rho}{1-\rho} -1}
\nonumber 
\end{eqnarray}
At the large distance limit, $\eta _{\rm F}^{\rm P}$ tends to unity,
that is the value obtained for perfect reflectors. At the small distance
limit, $\eta _{\rm F}^{\rm P}$ is found to vary as \cite{Lambrecht00} 
\begin{eqnarray}
L &\ll &\lambda _{\rm P} \quad \rightarrow \quad 
\eta _{\rm F}^{\rm P}\simeq \alpha \frac{L}{\lambda _{\rm P}}  \nonumber \\
\alpha  &=&\frac{30}{\pi ^{2}}\int_{0}^{\infty }{\rm d}K\ e^{-\frac{3K}{4}%
}\left( \frac{K^{2}}{\sqrt{\sinh \frac{K}{2}}}-\frac{K^{2}}{\sqrt{\cosh 
\frac{K}{2}}}\right)   \nonumber \\
&\simeq &1.193  \nonumber 
\end{eqnarray}


\begin{references}
\bibitem{Casimir48}  H.B.G. Casimir, {\it Proc. Kon. Nederl. Akad. Wet.}   
{\bf 51} 793 (1948)

\bibitem{Deriagin57}  B.V. Deriagin and I.I. Abrikosova, {\it Soviet Physics
JETP} {\bf 3} 819 (1957)

\bibitem{Sparnaay}  M.J. Sparnaay, {\it Physica} {\bf XXIV} 751 (1958); W.
Black, J.G.V. De Jongh, J.Th.G. Overbeek and M.J. Sparnaay, {\it %
Transactions of the Faraday Society} {\bf 56} 1597 (1960)

\bibitem{Tabor68}  D. Tabor and R.H.S. Winterton, {\it Nature} {\bf 219}
1120 (1968)

\bibitem{Sabisky73}  E.S. Sabisky and C.H. Anderson, {\it Phys. Rev.} 
{\bf A7} 790 (1973)

\bibitem{Lamoreaux97}  S.K. Lamoreaux, {\it Phys. Rev. Lett.} {\bf 78} 5
(1997); erratum in {\it Phys. Rev. Lett.} {\bf 81} 5475 (1998)

\bibitem{Mohideen98}  U. Mohideen and A. Roy, {\it Phys. Rev. Lett.} {\bf 81}
4549 (1998)

\bibitem{Roy99}  A. Roy, C. Lin and U. Mohideen, {\it Phys. Rev.} {\bf D 60}%
, 111101 (1999)

\bibitem{Fischbach}  E. Fischbach and C. Talmadge, {\it The Search for Non
Newtonian Gravity} (AIP Press/Springer Verlag, 1998) and references therein;
E. Fischbach and D.E. Krause, {\it Phys. Rev. Lett.} {\bf 82} 4753 (1999)

\bibitem{Carugno97}  G. Carugno, Z. Fontana, R. Onofrio and C. Rizzo, {\it %
Phys. Rev.} {\bf D55} 6591 (1997)

\bibitem{Bordag99}  M. Bordag, B. Geyer, G.L. Klimchitskaya, and V.M.
Mostepanenko, {\it Phys. Rev.} {\bf D60} 055004 (1999)

\bibitem{Lifshitz}  E.M. Lifshitz, {\it Sov. Phys. JETP} {\bf 2} 73 (1956);
E.M. Lifshitz and L.P. Pitaevskii, {\it Landau and Lifshitz Course of
Theoretical Physics: Statistical Physics Part 2} ch VIII
(Butterworth-Heinemann, 1980)

\bibitem{Mehra67}  J. Mehra, {\it Physica} {\bf 57} 147 (1967)

\bibitem{Brown69}  L.S. Brown and G.J. Maclay, {\it Phys. Rev.} {\bf 184}
1272 (1969)

\bibitem{Schwinger78}  J. Schwinger, L.L. de Raad Jr., and K.A. Milton, {\it %
Annals of Physics} {\bf 115} 1 (1978)

\bibitem{Lambrecht00}  A. Lambrecht and S. Reynaud, {\it Eur. Phys. J.}    
{\bf D}, to appear; quant-ph/9907105

\bibitem{Deriagin68}  B.V. Deriagin, I.I. Abrikosova and E.M. Lifshitz, {\it %
Quart. Rev.} {\bf 10}, 295 (1968)

\bibitem{Blocki77}  J. Blocki, J. Randrup, W.J. Swiatecki and C.F. Tsang, 
{\it Ann. Physics} {\bf 105}, 427 (1977)

\bibitem{Mostepanenko85}  V.M. Mostepanenko and N.N. Trunov {\it Sov. J.
Nucl. Phys.} {\bf 42} 812 (1985)

\bibitem{Bezerra97}  V.B. Bezerra, G.L. Klimchitskaya and C. Romero, {\it %
Mod. Phys. Lett.} {\bf 12}, 2613 (1997)

\bibitem{Klimchitskaya99}  G.L. Klimchitskaya, A. Roy, U. Mohideen and V.M.
Mostepanenko, {\it Phys. Rev.} {\bf A60} 3487 (1999)

\bibitem{Jaekel91}  M.T. Jaekel and S. Reynaud, {\it J. Physique} {\bf I-1}
1395 (1991)

\bibitem{Schulz57}  L.G. Schulz, {\it Phil. Mag. Suppl.} {\bf 6} 102 (1957)

\bibitem{Ehrenreich62}  H. Ehrenreich and H.R. Philipp, {\it Phys. Rev.}  
{\bf 128} 1622 (1962)
\end{references}
\end{document}